\documentclass[aps,prl,reprint,superscriptaddress,amsmath,amssymb,preprintnumbers.longbibliography]{revtex4-1}
\usepackage{graphicx}
\usepackage[colorlinks=true,linkcolor=blue,citecolor=blue, urlcolor=blue]{hyperref}   
\usepackage[capitalise]{cleveref}

\newcommand{\fakesection}[1]{%
  \par\refstepcounter{section}
  \sectionmark{#1}
  \addcontentsline{toc}{section}{\protect\numberline{\thesection}#1}
  \textit{#1.---}}

\begin{document}

\title{Hydrodynamic attractor in periodically driven ultracold quantum gases}

\author{Aleksas Mazeliauskas}
\email{a.mazeliauskas@thphys.uni-heidelberg.de}
\affiliation{Institute for Theoretical Physics, University of Heidelberg, 69120 Heidelberg, Germany}

\author{Tilman Enss}
\affiliation{Institute for Theoretical Physics, University of Heidelberg, 69120 Heidelberg, Germany}
\date{\today}

\begin{abstract}
Hydrodynamic attractors characterize hydrodynamic-like evolution in strongly interacting systems, independent of initial conditions or microscopic details, outside the conventional hydrodynamic regime.
  They explain why hydrodynamic models apply to high-energy nuclear collisions, but so far have only been explored for monotonic expansion, such as Bjorken flow. 
  We demonstrate that a system undergoing periodic expansion and contraction exhibits a novel cyclic attractor behavior. We employ Müller-Israel-Stewart theory to a driven ultracold Fermi gas to predict the shape of the attractor, which does not converge to Navier-Stokes dynamics at late times. This phenomenon can be measured in real time in ultracold quantum gases with externally modulated scattering length, offering a new avenue for the experimental discovery of hydrodynamic attractors.
\end{abstract}

\maketitle

\fakesection{Introduction}The study of how physical systems approach equilibrium is a recurring question in many branches of physics, from cosmology~\cite{Bassett:2005xm,Amin:2014eta} to ultracold atoms~\cite{Langen:2016vdb,Marino:2022eiw,Mikheev:2023juq} to high-energy nuclear collisions~\cite{Busza:2018rrf,Schlichting:2019abc,Berges:2020fwq}. 
In the context of QCD matter thermalization, an essential role is played by nonthermal~\cite{Berges:2008wm} and hydrodynamic~\cite{Heller:2015dha} attractors characterized by the rapid loss of initial state information.
While viscous (Navier-Stokes) hydrodynamics is the universal long distance and long time description near equilibrium~\cite{landau2013fluid}, the \emph{hydrodynamic attractor} refers to the hydrodynamic-like behavior when perturbations from equilibrium are still significant. The remarkable success of hydrodynamic models in describing the expansion of Quark Gluon Plasma created in high-energy nuclear collisions is credited to such rapid hydrodynamization~\cite{Busza:2018rrf,Schlichting:2019abc,Berges:2020fwq}.

Hydrodynamic attractors have been studied in many theoretical contexts: holography~\cite{Heller:2011ju,Heller:2015dha,Buchel:2016cbj,Romatschke:2017vte,Kurkela:2019set,Grozdanov:2019kge,Mitra:2024zfy}, QCD Kinetic Theory~\cite{Kurkela:2018vqr,Kurkela:2018wud,Kurkela:2018xxd,Kurkela:2018oqw,Almaalol:2020rnu,Boguslavski:2023jvg}, Relativistic Boltzmann Transport~\cite{Nugara:2023eku,Nugara:2024net}, Boltzmann RTA~\cite{Heller:2016rtz,Strickland:2017kux,Blaizot:2017ucy,Strickland:2018ayk,Blaizot:2019scw,Behtash:2019txb,Dash:2020zqx,Aniceto:2024pyc,Frasca:2024ege}, and generalized hydrodynamic theories~\cite{Romatschke:2017acs,Strickland:2017kux,Behtash:2017wqg,Rocha:2023hts,Chen:2024pez,Buza:2024jxe,Chen:2024grb}; see also review articles~\cite{Florkowski:2017olj,Romatschke:2017ejr,Soloviev:2021lhs,Jankowski:2023fdz}.
Hydrodynamic attractors have a rich mathematical structure based on transseries~\cite{Heller:2018qvh,Aniceto:2018bis,Aniceto:2024pyc}, and can be viewed more generally as dynamical dimensionality reduction of
phase space~\cite{Heller:2020anv}.
 Recently, a proposal to measure attractor behavior with ultracold atoms at strong interaction was presented in Ref.~\cite{fujii2024}. 
However, so far attractor behavior has been studied only in monotonically expanding or driven systems. In this Letter, we present the first study of hydrodynamic attractors in \emph{periodically} driven systems, which are especially relevant for ultracold atom experiments.

Hydrodynamic attractor behavior was first discovered in modeling the equilibration of high-temperature QCD matter in high-energy heavy-ion collisions. Right after the two nuclei pass each other, the system expands rapidly along the beam axis~\cite{Bjorken:1982qr}. The expansion rate is $\partial_\mu u^\mu = 1/\tau$, where $u^\mu$ is a 4-velocity and $\tau$ is the proper time after the collision. High-temperature QCD matter is approximately conformal, and Navier-Stokes hydrodynamics predicts that the expansion induces an anisotropy in the diagonal elements of the energy-momentum tensor proportional to the shear viscosity $\eta$~\cite{Romatschke:2017ejr},
\begin{equation}
\frac{T_\text{NS}^{xx}-P_\text{eq}}{P_\text{eq}} = \frac{2}{3} \frac{\eta}{\tau P_\text{eq}} \,.\label{eq:ani}
\end{equation}
The deviation from the equilibrium pressure $P_\text{eq}$ is controlled by the dimensionless time variable $\tilde w = \tau P_\text{eq}/(\pi \eta)={\tau T}/{(4\pi \eta/s)}$~\cite{Heller:2016rtz}, which is related to temperature $T$ and the celebrated shear viscosity to entropy ratio $\eta/s$~\cite{Policastro:2001yc}. For $\tilde w\to \infty$ the viscous corrections vanish and the system becomes isotropic. However, at early times $\tilde w\to 0$, the viscous corrections $1/\tilde w$ in \cref{eq:ani} are large and Navier-Stokes hydrodynamics becomes invalid.

The early-time equilibration dynamics have been studied in many microscopic theories~\cite{Florkowski:2017olj,Romatschke:2017ejr,Soloviev:2021lhs,Jankowski:2023fdz}. These studies reveal that the full pressure anisotropy $({T^{xx}-P_\text{eq}})/{P_\text{eq}}$ 
converges to the Navier-Stokes expectation already for short times $\tilde w =\mathcal{O}(1)$ when viscous corrections are still large.
Furthermore, for different initial conditions the pressure anisotropy relaxes to a common trajectory even before reaching the Navier-Stokes expectation. 
This emergent hydrodynamic-like behavior long before reaching equilibrium is known as the hydrodynamic attractor.

\begin{figure*}
\includegraphics[width=0.9\textwidth]{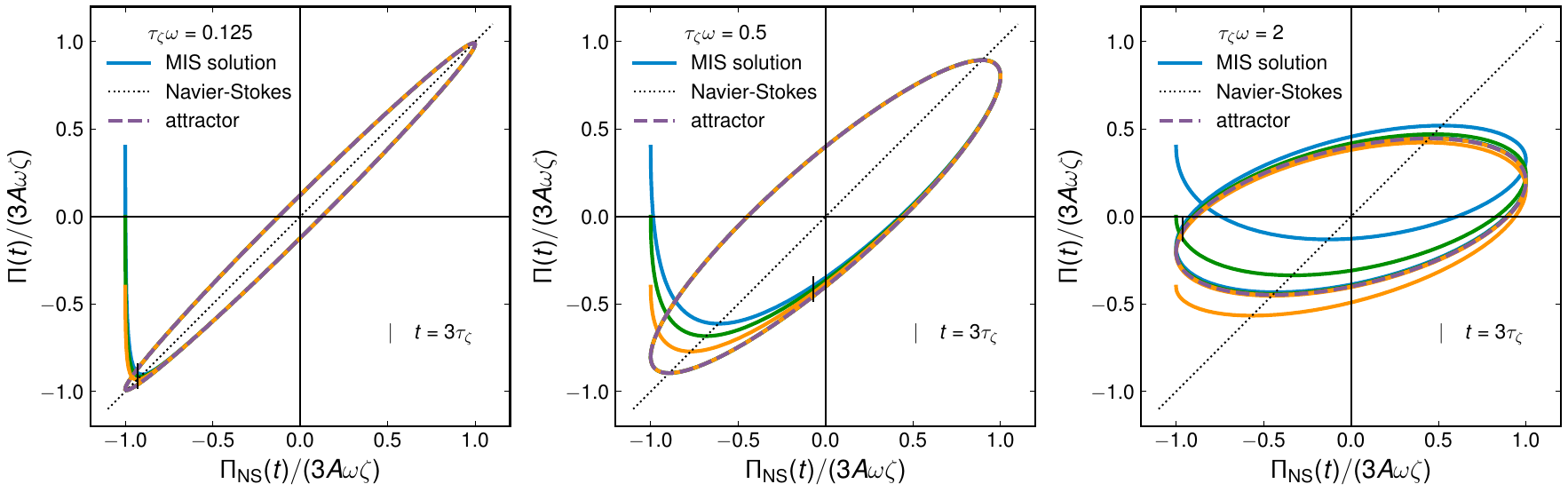}
\caption{M\"uller-Israel-Stewart hydrodynamics: linear attractor solution for oscillating expansion at different drive frequencies $\omega$. The full dynamical bulk pressure $\Pi(t) \equiv\tfrac{1}{3}T^{i}_{\phantom{i}i}-P_\text{eq}$ is plotted vs its Navier-Stokes approximation $\Pi_\text{NS}(t)$, such that deviations from Navier-Stokes appear outside the diagonal (dotted line). Different initial conditions (colors) converge toward an asymptotic attractor solution (dashed ellipse) near $t=3\tau_\zeta$, as marked by the vertical bar. For increasing drive frequency (left to right panels) the elliptic shape of the attractor widens; this signifies a growing deviation from Navier-Stokes. 
\label{fig:cycle}}
\end{figure*}

In this paper, we show that systems undergoing periodic expansion and contraction never relax to a Navier-Stokes limit. Instead, they settle for a new type of cyclic hydrodynamic attractor. For such systems, the expected deviation from equilibrium, \cref{eq:ani}, vanishes at certain turning points, hence the scaled time $\tilde w$ is not a meaningful time variable. Instead, we propose to plot the actual anisotropy $({T^{xx}-P_\text{eq}})/{P_\text{eq}}$ versus the Navier-Stokes expectation $({T_\text{NS}^{xx}-P_\text{eq}})/{P_\text{eq}}$. Deviations from the diagonal $y=x$ line serve as a clear indicator of departures from Navier-Stokes hydrodynamics.

Ultracold atom experiments offer the opportunity to measure the full time evolution of a strongly correlated quantum fluid and directly test where the attractor description is valid. Compared to previous approaches with monotonic expansion, cyclic attractors in periodically driven systems have the crucial advantage that attractor behavior is seen at longer observation times and does not require fine
time resolution. 
Whereas for Bjorken expansion the drive amplitude and effective “frequency” are intrinsically coupled, in periodically driven systems they can be tuned independently. 
Therefore, already with existing experimental techniques it is feasible to study attractor behaviour and improve our understanding of how and when systems start to behave hydrodynamically.

In recent years, the hydrodynamic properties of ultracold atomic gases have been extensively characterized, and the equation of state and transport coefficients have been obtained via collective mode excitations in harmonically trapped and uniform gases 
\cite{riedl2008, chiacchiera2009, bluhm2017, patel2020, li2022}. 
In particular, the damping of isotropic expansion dynamics is determined by the bulk viscosity $\zeta$. The three-dimensional, resonantly interacting Fermi gas constitutes an interesting special case: it is conformally invariant and has a vanishing bulk viscosity $\zeta$ \cite{son2007, nishida2007nonrel, enss2011, dusling2013, fujii2020}, as also confirmed experimentally \cite{elliott2014observation, wang2024}. In two dimensions, instead, scale invariance is broken by quantum fluctuations \cite{olshanii2010, hofmann2012, vogt2012, brewer2015, holten2018, peppler2018, murthy2019} that lead to nonzero damping.

Recent works demonstrate a new and more sensitive method to probe expansion dynamics in ultracold quantum gases: external modulation of the scattering length $a(t)$ is equivalent to fluid expansion and probes local dissipation even in a uniform system without actual fluid motion \cite{fujii2018, fujii2024}. Experimentally, continuous periodic modulation of the scattering length is an established technique that has been used to nonlinearly excite matter-wave jets \cite{clark2017} and density patterns \cite{zhang2020, fujii2024stable, liebster2025}, as well as the amplitude (``Higgs'') mode of the superfluid order parameter \cite{bayha2020}.
Here, we propose to use periodic modulation of the scattering length to probe cyclic attractors both in the linear and nonlinear regimes.

In the following, we first present how a new type of hydrodynamic attractors arises in periodically driven systems. In linear response these can be  described by M\"uller-Israel-Stewart hydrodynamics, which applies both to relativistic heavy-ion and nonrelativistic ultracold atom systems~\footnote{Neutron star mergers is another example of where MIS equation applies~\cite{Gavassino:2023xkt}}. Next, in order to study attractors in the nonlinear regime of large drive amplitude, we use relativistic kinetic theory for massive particles, which includes the ultracold atom case as the nonrelativistic limit.

\fakesection{M\"uller-Israel-Stewart theory}We consider a three-dimensional Fermi gas in an isotropic trap whose inverse scattering length $a^{-1}(t)$ is modulated periodically over time with a small amplitude~\footnote{Here we consider a finite equilibrium scattering parameter $a_0^{-1}$. We note that in ultracold gases it is also possible to modulate the scattering length around the scattering resonance at $a^{-1}_0=0$ because the bulk viscosity is suppressed near the resonance as $\zeta\sim a^{-2}$, yielding a finite dynamical response \cite{fujii2018, fujii2024}.}
\begin{equation}
a^{-1}(t)=a^{-1}_0(1+A \sin\omega t ), \quad A\ll 1.
\end{equation}
In ultracold Fermi gases this is achieved via Feshbach resonances \cite{chin2010feshbach}. As shown in Ref.~\cite{fujii2018}, the time variation of the scattering length leads to the same local dissipation as the homogeneous expansion of that gas with expansion rate $\theta =\partial_\mu u^\mu= 3 a(t) \partial_t a^{-1}(t)$. The linear response of the bulk pressure $\Pi(t) \equiv\frac{1}{3}T^{i}_{\phantom{i}i}-P_\text{eq}$ can be described by the Müller-Israel-Stewart (MIS) equation~\cite{Muller:1967zza,Israel:1979wp}
\begin{align}
\tau_\zeta \dot \Pi(t)  &=- \Pi(t)+ \Pi_\mathrm{NS}(t)\label{eq:MIS}
,\end{align}
where $\tau_\zeta$ is the relaxation time characterizing local dissipation, which has been computed microscopically for ultracold atoms \cite{dusling2013, nishida2019, enss2019bulk, hofmann2020, fujii2023}. Near a scattering resonance $\tau_\zeta$ is found to scale inversely with temperature, $\tau_\zeta \simeq 0.7\hbar/k_BT$ \cite{enss2019bulk, fujii2024}. The Navier-Stokes expectation $\Pi_\mathrm{NS}(t)=-\zeta[a(t)]\theta $ corresponds to very fast relaxation and serves as a driving term. The general solution of \cref{eq:MIS} for small-amplitude drive $\Pi_\mathrm{NS}(t)$ starting at $t=0$ is given by
\begin{equation}
\Pi(t) = e^{-t/\tau_\zeta}  \Pi(0) + \frac{1}{\tau_\zeta}\int_0^t dt'\, e^{(t'-t)/\tau_\zeta}  \Pi_\mathrm{NS}(t').\label{eq:fullPi}
\end{equation}

It is straightforward to show that the late-time limit of \cref{eq:fullPi} is
\begin{equation}
\Pi(t)=-3 A \omega \zeta \frac{\cos(\omega t-\phi )}{\sqrt{(\omega\tau_\zeta)^2+1}},\label{eq:latetime}
\end{equation}
where $\zeta=\zeta(a_0), \tau_\zeta=\tau_\zeta(T_0)$ and the phase shift is given by
$\tan \phi = \omega\tau_\zeta$. For nearly instantaneous  relaxation or very slow drives, $ \omega \tau_\zeta \ll 1$, we recover the Navier-Stokes limit $\Pi_\mathrm{NS}(t) = -3 A\omega \zeta \cos(\omega t)$. However, even for moderately slow drives the MIS attractor will never converge to the NS result.

\begin{figure*}
\includegraphics[width=0.9\textwidth]{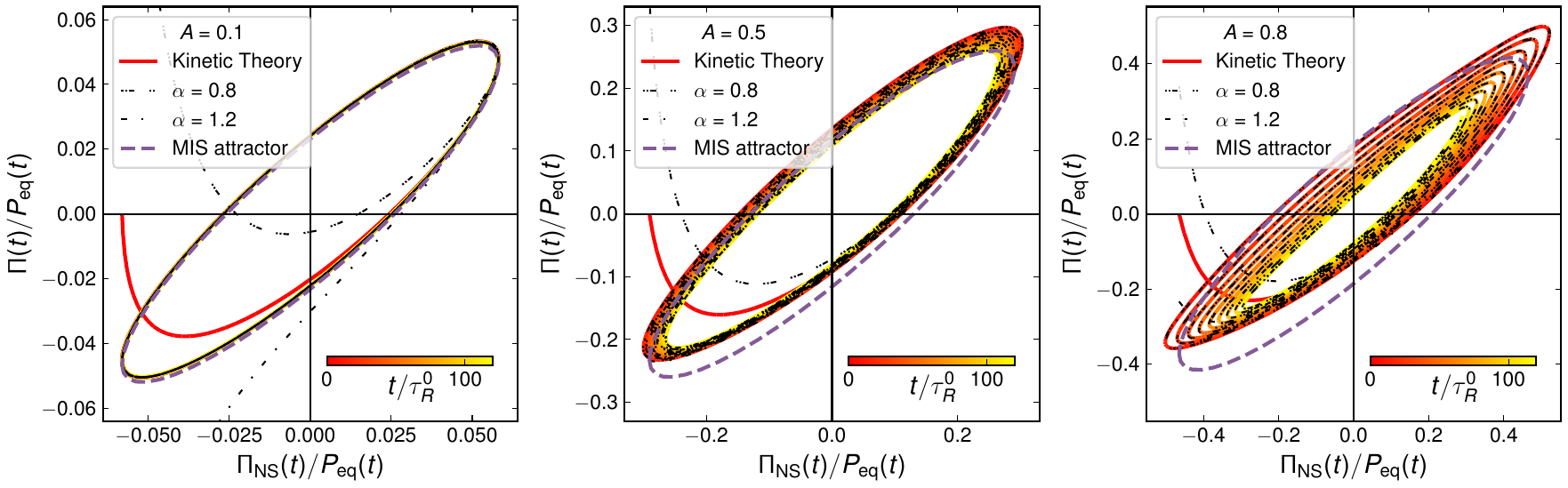}
\caption{Massive kinetic theory: nonlinear attractor solution for oscillating expansion at different drive amplitude $A$ and different initial conditions $\alpha$. At small amplitude (left panel) the full kinetic theory solution (solid line) agrees with the linear MIS attractor (dashed line). At larger amplitude (center and right panels) kinetic theory predicts a nonlinear attractor that deviates strongly from the MIS attractor. The asymmetric attractor shape arises because also the equation of state evolves with time, cf.\ \cref{fig:temperature}.
\label{fig:mKT}}
\end{figure*}

The bulk pressure has units of energy density. Therefore, it is convenient to normalize $\Pi$ by the equilibrium pressure $P_\text{eq}$ and by the bulk viscosity to $\tau_\zeta$ ratio normalized by pressure, $\zeta/(\tau_\zeta P_\text{eq})$, which are properties of the fluid and universal scaling functions of temperature $T/T_F$ and interaction $a_0$.  Their values from microscopic computations for strongly correlated Fermi gases are reported in Refs.~\cite{enss2019bulk, haussmann2007}. Finally, the linear response is proportional to the characteristic perturbation amplitude $3A\omega \tau_\zeta$. Therefore, we consider the ratio
\begin{equation}
\frac{\Pi}{P_\text{eq}} \left(\frac{\zeta}{P_\text{eq} \tau_\zeta}\right)^{-1} \left(3A \omega \tau_\zeta\right)^{-1} = \frac{\Pi}{3A \omega\zeta},
\end{equation}
which is the universal dimensionless bulk pressure. We plot the time evolution of $\Pi(t)$ against the expected Navier-Stokes evolution 
in \cref{fig:cycle} for different oscillation frequencies $\omega \tau_\zeta$ and initial values of $\Pi(0)$. We observe that the full solution approaches the late-time limit, \cref{eq:latetime}, already at time $t/\tau_\zeta=3$.
For slow drives $\omega \tau_\zeta \ll 1$, different initial conditions collapse onto the narrow ellipse along the Navier-Stokes limit (diagonal). The width of the ellipse can be defined as the extent $\Delta\Pi$ of the bulk pressure $\Pi$ where $\Pi_\text{NS}=0$, and is given by $\Delta\Pi/(3A\omega\zeta) = 2\omega\tau_\zeta/[1+(\omega\tau_\zeta)^2]\leq1$. The maximum width is reached when the drive frequency $\omega=\tau_\zeta^{-1}$ equals the relaxation rate. For higher values of the drive frequency, the phase shift $\phi$ between Navier-Stokes and actual bulk pressure increases, but the width decreases. This is because at large drive frequencies, the system is not able to follow and respond quickly enough. It is a crucial experimental advantage of periodic drives over monotonic expansion that $\tau_\zeta$ can be determined not only from the short-time transient $e^{-t/\tau_\zeta}$ (nonhydrodynamic mode) but also from the phase shift $\phi$ at long times.

\fakesection{Massive kinetic theory}In order to study hydrodynamic attractors in the nonlinear regime of large drive amplitudes, we employ the relativistic kinetic theory of massive particles in relaxation time approximation. One could also solve the MIS equations in the nonlinear regime, but the kinetic approach guarantees that the net pressure $P_\text{eq}+\Pi$ remains positive even for large amplitudes.

We generalize Refs.~\cite{Florkowski:2014sfa,Keegan:2015avk} to a time-dependent metric $ds^2 = dt^2- b(t)^2(dx^2+dy^2+dz^2)$ with scale factor $b(t)$, which is equivalent to the homogeneous and isotropic expansion of gas with rate $\theta = 3 \dot b/b$. This isotropic expansion is equivalent to the change of scattering length discussed above \cite{fujii2018} and a convenient way to compute the bulk viscosity \cite{son2007}. In spherical coordinates the phase-space distribution function obeys the following Boltzmann equation~\cite{cercignani2002relativistic,Du:2021fok}
\begin{equation}
\left[\partial_t - 2 \frac{\dot b}{b}p \frac{\partial }{\partial p}\right]f(t,p) = - \frac{f(t,p)-f_\text{eq}(p,T(t))}{\tau_R(t)},\label{eq:Boltzman}
\end{equation}
where $\tau_R$ is the relaxation time.
As a realistic model for $\tau_R$ we use the bulk relaxation time $\tau_\zeta$ obtained for strongly interacting ultracold atoms from microscopic quantum transport calculations~\cite{enss2019bulk, fujii2024}. Accordingly, we set $\tau_R$ inversely proportional to temperature, $\tau_R(t) = \tau_R^0 T_0/T(t)$.

The equilibrium distribution $f_\text{eq}(p,T) = e^{-\sqrt{m^2+p^2 b^2(t)}/T}$ depends on temperature $T$. Out of equilibrium the temperature changes with time, and one can define an instantaneous effective temperature $T(t)$ by the condition that the instantaneous energy density matches the equilibrium energy density at this instantaneous temperature, $e(t)=e_\text{eq}(T(t),m)$, where
\begin{align}
e(t)&=\int\frac{d^3 p}{(2\pi)^3}b(t)^3 \sqrt{m^2+b(t)^2p^2}  f(t,p),\label{eq:et}\\
e_\text{eq}(T,m)&= \frac{3 T^4}{\pi^2}
\left( \frac{m^3}{6T^3}K_1\left(\frac{m}{T}\right) + \frac{m^2}{2T^2} K_2\left(\frac{m}{T}\right)\right).\label{eq:eTh}
\end{align}
Analogously, the particle number density can be fixed by introducing a chemical potential $\mu$.

The implicit general solution of the Boltzmann \cref{eq:Boltzman}~\cite{Florkowski:2013lya,Florkowski:2014sfa,Du:2021fok} is
\begin{align}
    f(t, p) &=  e^{-\int_0^t dt'/\tau_R(t')} f_0(p b^2(t))\nonumber \\
&+\int_{0}^t \frac{dt'}{\tau_R(t')}  e^{-\int_{t'}^t dt''/\tau_R(t'')} f_\text{eq}\left(p \frac{b^2(t)}{b^2(t')}, T(t')\right),\label{eq:KTsolution}
\end{align}
where one has to determine self-consistently the effective temperature evolution $T(t)$. This can be done by substituting the distribution, \cref{eq:KTsolution}, into the energy density integral, \cref{eq:et}, and equating it to the equilibrium energy density, \cref{eq:eTh}. The resulting integral equation can be solved iteratively for $T(t)$ and then inserted into the solution for the distribution, \cref{eq:KTsolution}  (see End Matter for details).  At $t=0$ we consider a one parameter family of distribution functions $f_0(p) =N_\alpha f_\text{eq}(\alpha p, T_0)$ where $N_\alpha$ is chosen such that that initial energy density remains constant, i.e., $T(0)=T_0$.

We solve \cref{eq:Boltzman} for periodic expansion and contraction described by $b(t) = 1+A \sin \omega t$. In particular, we compute the time evolution of  bulk pressure $\Pi(t) \equiv\frac{1}{3}T^{i}_{\phantom{i}i}-P_\text{eq}$.
The Navier-Stokes expectation is $\Pi_\mathrm{NS}(t) = -3A\omega \zeta \cos( \omega t)/(1+A \sin (\omega t))$. For massive Boltzmann particles, the ratio of bulk viscosity to pressure $\zeta/(\tau_R P_\text{eq})$ is a monotonically increasing function of $m/T$ between zero and $2/3$, as reported in Fig.~\ref{fig:zeta} in the End Matter~\cite{Florkowski:2014sfa}.
In each panel of \cref{fig:mKT} we show the ratio $\Pi(t)/P_\text{eq}$ for different amplitudes $A=0.1,0.5,0.8$. In each panel we show evolutions for three different initial conditions: $\alpha=1.0$ (solid line), $\alpha=0.8$, and $\alpha=1.2$ (black dot-dashed lines). For concreteness, we consider $m/T=10$, $\omega\tau_R=0.5$, and $\zeta/(\tau_R P_\text{eq})=0.39$ at $t=0$.
 For small amplitude $A=0.1$ (left panel), the full kinetic solution approaches the linear-response MIS attractor shown by dashed line. For a larger amplitude $A=0.5$ (central panel), the kinetic theory attractor deviates from MIS attractor. In the right-most panel of \cref{fig:mKT} we show the kinetic solution in a strongly nonlinear regime $A=0.8$. 

\begin{figure}
    \centering
  \quad  \includegraphics[width=0.9\linewidth]{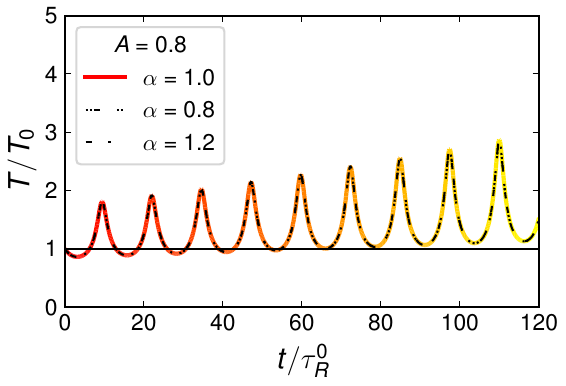}
    \includegraphics[width=0.95\linewidth]{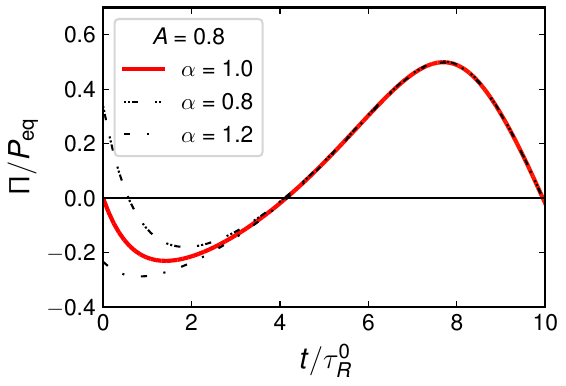}
    \caption{(top) Temperature evolution for a strong periodic drive. The temperature oscillation has large amplitude, deviates from the sinusoidal shape of the drive, and the peak height grows with each cycle due to dissipative heating. This temperature dependence leads to the asymmetric orbits of the attractor in Fig.~\ref{fig:mKT}(c), which do not close after one period.
    (bottom) Bulk pressure evolution for different initial conditions and a strong periodic drive.
    }    \label{fig:temperature}
\end{figure}

The nonlinear attractor from kinetic theory is no longer strictly periodic and its trajectory does not close. Instead, it
drifts in time, which we associate with the time-evolving equation of state. In \cref{fig:temperature}(top) we show the temperature evolution in the strongly nonlinear regime $A=0.8$. Due to entropy production, the system is heating up and the average temperature increases with time. This modifies the average $\zeta/(\tau_R P_\text{eq})$ and $P_\text{eq}$ values. The cycle averaged energy grows in time quadratically with drive amplitude $A$ as $\langle \dot e \rangle \propto \zeta A^2 \omega^2$ \cite{fujii2018}, which was neglected in linear response. Nevertheless, the nonlinear attractor is a universal attractor because different initial conditions quickly converge toward the same curve. This is seen in \cref{fig:mKT}, where the evolution for different initial values for the bulk pressure (black dash-dotted lines) falls onto the same attractor curve. For clarity, \cref{fig:temperature}(bottom) focuses on the short-time evolution, which demonstrates that also in the strongly nonlinear regime convergence onto the attractor occurs within $t\simeq 3\tau_R^0$.
Finally, we note that the attractor drift in \cref{fig:mKT} arises predominantly from amplitude evolution rather than frequency variation. Hence, the accurate experimental determination of the phase difference between the Navier–Stokes and attractor responses (set by the relaxation time) remains possible.

We expect that the kinetic theory predictions for attractors in the linear and nonlinear regimes will also hold in the non-relativistic limit  ($m/T\gg1$) for a nondegenerate atomic gas. One will need to introduce chemical potential and include the dependence of the bulk susceptibility $\zeta/\tau_R P_\text{eq}\sim z(\lambda/a)^2$ on thermal length $\lambda$, scattering length $a$ and fugacity $z$ near resonance \cite{dusling2013, enss2019bulk}. The results could then be compared with experiments at both small and large drive amplitude. We leave these extensions for future work.

\fakesection{Conclusions}
We have shown how the response to periodic drives realizes a new form of cyclic hydrodynamic attractor beyond the paradigm of monotonic expansion. These attractors are solutions of MIS hydrodynamics or kinetic theory and differ from the Navier-Stokes prediction even at late times and for slow drives. This is a crucial advantage for their experimental identification, where a long-time measurement can yield very precise results. Such measurements are already feasible with ultracold atoms, where the out-of-equilibrium equation of state has be measured with high time resolution \cite{bardon2014, luciuk2017, xie2025}.
The observation of nontrivial phase-space trajectories that are independent of initial conditions (see \cref{fig:mKT}) would clearly indicate attractor behavior.
This would manifest not only in the linear regime but also in the nonlinear regime under strong driving conditions.

A key advantage of periodic drives is the ability to control the drive frequency $\omega$ and amplitude $A$ independently. In Bjorken expansion, both MIS and RTA kinetic theory exhibit early-time (free-streaming) attractor behavior driven by the large expansion rate $\partial_\mu u^\mu = \tau^{-1} \gg \tau_\zeta^{-1}$~\cite{Kurkela:2019set,Blaizot:2019scw}. In contrast, periodic drives allow the free-streaming regime $\omega \tau_\zeta \gg 1$ to be probed at both small and large amplitudes, whereas in the opposite limit $\omega \tau_\zeta \ll 1$ with large amplitudes, one can access genuinely nonlinear hydrodynamic attractors.

Our work motivates the study of attractors for generic drives beyond the sinusoidal shape. These can be studied in the MIS and kinetic theory formulations presented in this Letter. Moreover, cyclic attractors can also be studied in holographic theories of strongly interacting systems, which is a promising route to understand them from the transseries point of view~\cite{Buchel:2016cbj}.

In summary, the experimental observation of hydrodynamic attractor phenomena in ultracold atom systems would be a remarkable validation of a theoretical concept originally developed in the study of high-energy nuclear collisions. Such a discovery would further deepen the phenomenological connections between these two research communities~\cite{Floerchinger:2021ygn,Brandstetter:2023jsy,Berges:2025izb}.

\fakesection{Acknowledgements}
We thank 
Jürgen Berges,
Wojciech Florkowski, 
Keisuke Fujii, 
Oscar Garcia-Montero, 
Michal Heller,
Aleksi Kurkela, 
Thimo Preis, 
Sören Schlichting, 
Joseph Thywissen, 
and 
Urs Wiedemann  
for useful discussions.
This work is supported by the DFG through Emmy Noether Programme (project number 496831614) (A.M.), through CRC 1225 ISOQUANT (project number 27381115) (T.E., A.M.) and through Germany's Excellence Strategy EXC 2181/1-390900948 (the Heidelberg STRUCTURES Excellence Cluster) (T.E.). We thank the ExtreMe Matter Institute EMMI at GSI, Darmstadt, for support in the framework of an EMMI Rapid Reaction Task Force meeting ``Deciphering many-body dynamics in mesoscopic quantum gases'', during which this work has been initiated. We acknowledge the use of ChatGPT (\href{http://chat.openai.com}{chat.openai.com}) to improve the readability of certain sentences in the paper.

\appendix
\section{End Matter: Massive kinetic theory}\label{app:KT}

We consider a homogeneous and isotropically expanding system  (Hubble expansion) of massive particles in a  Friedmann-Lemaître-Robertson-Walker (FLRW) metric (see \cite{Bazow:2015dha,Du:2021fok} for Hubble expansion of massless particles, \cite{Florkowski:2014sfa} for massive particles in Bjorken expansion, and \cite{Keegan:2015avk} for longitudinal expansion at weak and strong couplings)
\begin{equation}
ds^2 = dt^2- b(t)^2(dx^2+dy^2+dz^2).\label{eq:metric}
\end{equation}
The nonzero Christoffel symbols are $\Gamma^{i}_{0j}= \delta^i_j \dot b/b$, and $\Gamma^{0}_{ij}= b\dot b \delta_{ij}$, where $i,j$ are spatial indices. For momentum integrals, we need the integration measure $\sqrt{-g}=b^3$, where $g$ is the determinant of the metric. We assume that in the coordinates of \cref{eq:metric}, the fluid is static, i.e., $u^{\mu}=(1,0,0,0)$, but the expansion rate $\theta = D_\mu u^\mu= 3 \dot b/b$ is nonzero, where $D_\mu$ is the covariant derivative.

For homogeneous system the single-particle distribution function $f(t,\vec p)$ obeys the relativistic Boltzmann equation~\cite{cercignani2002relativistic,Du:2021fok}
\begin{equation}
\left[p^\mu \partial_\mu  - \Gamma^i_{\alpha \beta}p^\alpha p^\beta \frac{\partial }{\partial p^i}\right]f(t, \vec p) = - C[f],
\end{equation}
where $p^0 = \sqrt{m^2+|\vec p|^2 b^2}$ and $\vec p \equiv (p^1, p^2,p^3)$. For the FLRW metric, \cref{eq:metric}, we can simplify the Boltzmann equation to
\begin{equation}
\left[\partial_t - 2 \frac{\dot b}{b}p \frac{\partial }{\partial p}\right]f(t,p) = - \frac{1}{p^0} C[f].
\end{equation}
This can be solved by employing a scaled momentum $ \tilde p = p b^2(t)/b^2(t_0)$ and defining $F(t, \tilde p ) = F(t, p b^2(t)/b^2(t_0) )\equiv f(t,p)$.
In these coordinates the Boltzmann equation reads
\begin{equation}
\left.\partial_t F(t,\tilde p)\right|_{\tilde p} = -\frac{1}{p^0} C[F]\label{eq:F}.
\end{equation}
For collisionless systems, the solution is $f(t,p)=F(t_0, \tilde p ) = f(t_0, p b^2(t)/b^2(t_0) )$.

We consider kinetic theory in Relaxation Time Approximation (RTA), for which the collision kernel is given by 
\begin{equation}
C[f] =\frac{-p^\mu u_\mu \left(f(t, \vec p)-f_\text{eq}(t,\vec p )\right)}{\tau_R(t)}.
\end{equation}
For computational simplicity, we will consider classical Boltzmann particles for which the equilibrium distribution is determined by temperature $T$ as
\begin{equation}
f_\text{eq}(p, T) =e^{-\sqrt{m^2+b^2 p^2}/T}.
\end{equation}
For the collision kernel to satisfy energy conservation, we require that the instantaneous energy densities of $f(t, \vec p)$ and $f_\text{eq}(t,\vec p )$ are equal: $e(t)=e_\text{eq}(T(t),m)$, where  (degeneracy $\nu=1$)
\begin{align}
e(t)&=\int\frac{d^3 p}{(2\pi)^3}b(t)^3 \sqrt{m^2+b(t)^2p^2}  f(t,p),\\
e_\text{eq}(T,m)&= \int\frac{d^3 p}{(2\pi)^3}b(t)^3 \sqrt{m^2+b(t)^2p^2}  f_\text{eq}(p, T)\nonumber \\
=&\frac{3 T^4}{\pi^2}
\left( \frac{m^3}{6T^3}K_1\left(\frac{m}{T}\right) + \frac{m^2}{2T^2} K_2\left(\frac{m}{T}\right)\right),
\end{align}
and the time dependence of $T(t)$ has to be determined self-consistently.

\begin{figure*}
    \centering
    \includegraphics[width=\linewidth]{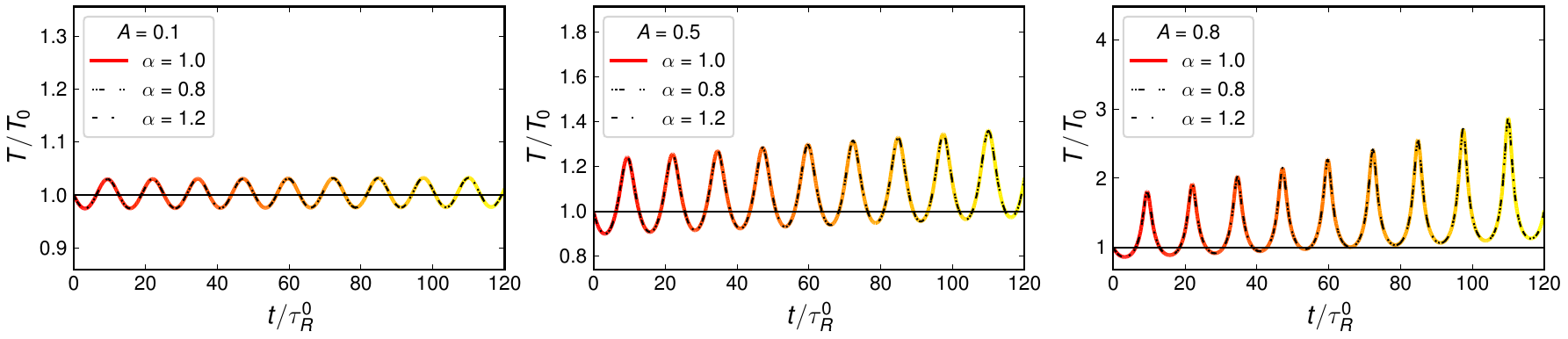}
    \caption{Temperature evolution for different periodic drives in massive kinetic theory. For larger drive amplitude $A$, the temperature oscillations grow in amplitude, they become less sinusoidal, and the peak height grows from one cycle to the next. However, the oscillation phase for different drive amplitudes is approximately the same even after nine periods.}
    \label{fig:Tevolution}
\end{figure*}

For the RTA collision kernel, the Boltzmann equation \cref{eq:F} can be written as
\begin{equation}
\partial_t F(t,\tilde p) =  \frac{F_\text{eq}(T(t),\tilde p)-F(t,\tilde p)}{\tau_R(t)},
\end{equation}
which has the solution (we take $t_0=0$ and $b(t_0)=1$)
\begin{align}
    F(t,\tilde p) &=   F(0,\tilde p) e^{-\int_0^t \frac{dt'}{\tau_R(t')}}\nonumber\\
    &+\int_{0}^t \frac{dt'}{\tau_R(t')}  e^{-\int_{t'}^t \frac{dt''}{\tau_R(t'')}} F_\text{eq}(\tilde p, T(t'))
\end{align}
or, alternatively (cf.~\cite{Du:2021fok}),
\begin{align}
   & f(t, p) =  e^{-\int_0^t \frac{dt'}{\tau_R(t')}} f(0,p b^2(t))\nonumber \\
&+\int_{0}^t \frac{dt'}{\tau_R(t')}  e^{-\int_{t'}^t \frac{dt''}{\tau_R(t'')}} f_\text{eq}\left(p \frac{b^2(t)}{b^2(t')}, T(t')\right).
\end{align}
We take the initial distribution to be a deformed equilibrium distribution, i.e., $f(0,p) =N_\alpha f_\text{eq}(\alpha p, T_0)$, where $\alpha>0$ is some positive constant. We choose the normalization $N_\alpha$, such that the initial energy density is fixed to $e_\text{eq}(T_0,m)$, i.e., $T(0)=T_0$.
Then
\begin{align}
    &f(t, p) =N_\alpha  e^{-\sqrt{m^2+\alpha^2 p^2 b^4(t)}/T_0}e^{-\int_0^t \frac{dt'}{\tau_R(t')}}\nonumber\\
    &+\int_{0}^t \frac{dt'}{\tau_R(t')}  e^{-\int_{t'}^t \frac{dt''}{\tau_R(t'')}} e^{-\sqrt{m^2+p^2 b^4(t)/b^2(t')}/T(t')}\label{eq:KTeq1}
\end{align}
In order to solve \cref{eq:KTeq1} we need to determine the instantaneous equilibrium temperature $T(t)$. Performing the energy density integral of \cref{eq:KTeq1} we obtain
\begin{align}
    \frac{3 T(t)^4}{\pi^2} &
\left( \frac{m^3}{6T(t)^3}K_1\left(\frac{m}{T(t)}\right) + \frac{m^2}{2T(t)^2} K_2\left(\frac{m}{T(t)}\right)\right) \nonumber \\
=&\, T_0^4 N_\alpha I\left(\frac{m}{T_0}, \alpha b(t)\right)e^{-\int_0^t \frac{dt'}{\tau_R(t')}}\nonumber\\
    &+\int_{0}^t \frac{dt'}{\tau_R(t')}  e^{-\int_{t'}^t \frac{dt''}{\tau_R(t'')}} T^4(t') I\left(\frac{m}{T(t')}, \frac{b(t)}{b(t')}\right),\label{eq:Teq}
\end{align}
where we used that $e(t)=e_\text{eq}(T(t))$ and defined a dimensionless integral
\begin{align}
I(y,z) &=\int_0^\infty\frac{4 \pi x^2 dx}{(2\pi)^3}\sqrt{y^2+x^2}  e^{-\sqrt{y^2+z^2 x^2}}.
\end{align}
Note that $\tau_R$ is assumed to be inversely proportional to the temperature,  $\tau_R(t) = \tau_R^0 T_0/T(t)$.

For each new time step $t=n \delta t$, we take an initial guess of $T(n \delta t)=T((n-1) \delta t)$. We use cubic spline interpolation of $T(t)$ in the integrals on the right-hand side of \cref{eq:Teq} to compute the energy density. We then perform a root-finding algorithm to determine an updated value for $T(n \delta t)$. We perform this procedure until the temperature change falls below $10^{-8}$ and then proceed to the next time step. The \texttt{C++} implementation of the solver converges under a minute for typical settings: $t_\text{max}=60$, $N_t=1000$, $\tau_R=0.5$, $m=10$, $\omega=1$, $T_0=1$, $A=0.5$.
We show the temperature evolution for $A=0.1,0.5,0.8$ and $\alpha=1,0.8,1.2$ in \cref{fig:Tevolution}. For small amplitudes, the temperature is oscillating symmetrically around the equilibrium value. For larger amplitude values the oscillation is asymmetric and deviates from the sinusoidal drive. We also observe a drift in average temperature, which is due to entropy production during the cycle. 

\begin{figure}
    \centering
    \includegraphics[width=0.9\linewidth]{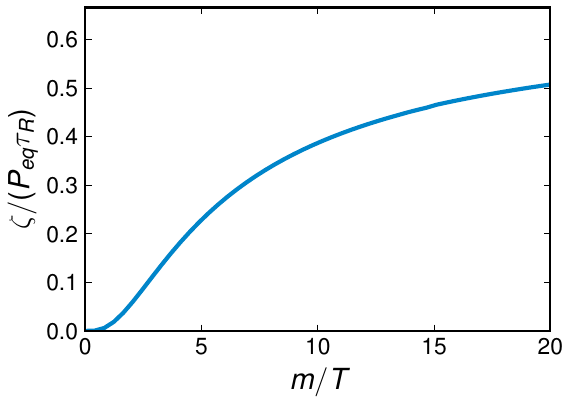}
    \caption{The bulk susceptibility $\zeta/P_\text{eq}\tau_R$ in massive kinetic theory as a function of $m/T$.}
    \label{fig:zeta}
\end{figure}

In the main text of the paper, \cref{fig:mKT}, we discuss the time evolution of the bulk pressure. In kinetic theory we define the bulk pressure as
\begin{align}
\Pi(t)&=\frac{1}{3} \int\frac{d^3 p}{(2\pi)^3} b^3\frac{p^2 b^2(t)}{\sqrt{m^2+p^2 b^2(t)}}  (f(t,p) - f_\text{eq}(t,p))\nonumber\\
&=\frac{1}{3}R_\text{eq}(T(t))-\frac{1}{3} \int\frac{d^3 p}{(2\pi)^3} b^3\frac{m^2}{\sqrt{m^2+p^2 b^2(t)}}  f(t,p)\label{eq:Pi}
\end{align}
where we used integration by parts and that $e(t)=e_\text{eq}(T(t))$. In addition, we defined the trace of the energy-momentum tensor (conformal scale breaking term) as
\begin{align}
R_\text{eq}(T)&= e-3P_\text{eq}\nonumber\\
&= \int\frac{d^3 p}{(2\pi)^3}b(t)^3 \frac{m^2}{\sqrt{m^2+b(t)^2p^2}}  f_\text{eq}(p, T)\nonumber \\
&=\frac{3 T^4}{\pi^2}
\left( \frac{m^3}{6T^3}K_1\left(\frac{m}{T}\right)\right).
\end{align}
Inserting the solution \cref{eq:KTeq1} into \cref{eq:Pi} we obtain
\begin{align}
\Pi(t)&=\frac{1}{3} R_\text{eq}(T(t))-T_0^4 N_\alpha J\left(\frac{m}{T_0}, \alpha b(t)\right)e^{-\int_0^t \frac{dt'}{\tau_R(t')}}\nonumber\\
    &-\int_{0}^t \frac{dt'}{\tau_R(t')}  e^{-\int_{t'}^t \frac{dt''}{\tau_R(t'')}} T^4(t') J\left(\frac{m}{T(t')}, \frac{b(t)}{b(t')}\right),
\end{align}
in terms of the integral
\begin{align}
J(y,z) &=\frac{1}{3}\int_0^\infty\frac{4 \pi x^2 dx}{(2\pi)^3}\frac{y^2}{\sqrt{y^2+x^2}}  e^{-\sqrt{y^2+z^2 x^2}}.
\end{align}
For completeness, we give the analytical form of the bulk susceptibility~\cite{Florkowski:2014sfa} as a function of $z=m/T$ and plot it in \cref{fig:zeta}: the analytical form reads
\begin{equation}
\frac{\zeta}{P_\text{eq}\tau_R} =  \frac{z^3}{9}\left(-\frac{K_2(z)}{3K_3(z)+zK_2(z)}+\frac{K_1(z)}{K_2(z)}-\frac{K_{i,1}(z)}{K_2(z)}\right)\label{eq:zeta_anal}
\end{equation}
where
\begin{equation}
K_{i,1}(z)=\frac{\pi}{2}[1-zK_0(z)L_{-1}(z)-zK_1(z)L_0(z)]
\end{equation}
and $K$ and $L$ are modified Bessel and Struve functions, respectively.

We note that the relation in \cref{eq:zeta_anal} is imposed by the structure of RTA kinetic equations. Therefore, $\tau_R$ and $\zeta$ cannot be chosen independently. In this work, we chose to $\tau_R\propto 1/T$ motivated by microscopic  quantum transport calculations~\cite{enss2019bulk, fujii2024}.
\bibliography{main,all}
\end{document}